Science
Education

# Do Inquiring Minds Have Positive Attitudes? The Science Education of Preservice Elementary Teachers

CATHERINE RIEGLE-CRUMB,[1] KARISMA MORTON,[1] CHELSEA MOORE,[2] ANTONIA CHIMONIDOU,[3] CYNTHIA LABRAKE,[4] SACHA KOPP[5]
[1]*Department of Curriculum and Instruction, STEM Education, University of Texas at Austin;* [2]*Department of Psychology, University of Massachusetts, Amherst;* [3]*UTeach Primary, College of Natural Sciences, University of Texas at Austin;* [4]*Department of Chemistry, University of Texas at Austin; and* [5]*College of Arts and Sciences, SUNY Stonybrook*

**ABSTRACT:** Owing to their potential impact on students' cognitive and noncognitive outcomes, the negative attitudes toward science held by many elementary teachers are a critical issue that needs to be addressed. This study focuses on the science education of preservice elementary teachers with the goal of improving their attitudes *before* they begin their professional lives as classroom teachers. Specifically, this study builds on a small body of research to examine whether exposure to inquiry-based science content courses that actively involve students in the collaborative process of learning and discovery can promote a positive change in attitudes toward science across several different dimensions. To examine this issue, surveys and administrative data were collected from over 200 students enrolled in the *Hands on Science* (HoS) program for preservice teachers at the University of Texas at Austin, as well as more than 200 students in a comparison group enrolled in traditional lecture-style classes. Quantitative analyses reveal that after participating in HoS courses, preservice teachers significantly increased their scores on scales measuring confidence, enjoyment, anxiety, and perceptions of relevance, while those in the comparison group experienced a decline in favorable attitudes to science. These patterns offer empirical support for the attitudinal benefits of inquiry-based instruction and have implications for the future learning opportunities available to students at all education levels.     © 2015 Wiley Periodicals, Inc. *Sci Ed* **99:**819–836, 2015

*Correspondence to*: Dr. Catherine Riegle-Crumb; e-mail: riegle@austin.utexas.edu

## INTRODUCTION

While the metaphor of science, technology, engineering, and mathematics (STEM) as a pipeline has been rightly criticized as too simplistic, it is nevertheless clear that students' early experiences in science classrooms shape their future achievement and interests (Xie & Shauman, 2003). With the goal of better understanding and ultimately improving elementary science education in the United States, researchers and policymakers have increased their attention toward teachers. A growing body of research now focuses on the science content knowledge of elementary science teachers, as the subject matter expertise that they possess has clear implications for what students learn (Diamond, Maerten-Rivera, Rohrer, & Lee, 2014; Heller, Daehler, Wong, Shinohara, & Miratrix, 2012; Kanter & Konstantopoulus, 2010; Sadler, Sonnert, Coyle, Cook-Smith, & Miller, 2013). Compared to secondary teachers, elementary teachers are much more likely to be trained as generalists and consequently less likely to have extensive content knowledge, and this pattern appears particularly pronounced for science (Haefner & Zembal-Saul, 2004). Clearly there are continued concerns about the need for teacher training programs and professional development to focus on increasing subject matter expertise.

Yet content knowledge is a necessary but insufficient characteristic of a successful teacher. Teachers' attitudes about the content they teach is another critical factor that has implications for classroom learning, and importantly, negative attitudes can exist independent of content-area expertise (Beilock, Gunderson, Ramirez, & Levine, 2010; Tosun, 2000). While there is comparatively less research on elementary teachers' attitudes toward science than math (Bursal & Paznokas, 2006), there is nevertheless evidence that many elementary teachers are not favorably inclined toward science. Such negative attitudes on the part of teachers can impact their students' attitudes toward science and can inhibit students' learning (Beilock et al., 2010; Jarrett, 1999; Ramey-Gassert, Shroyer, & Staver, 1996), creating a vicious cycle that must be interrupted. However, effectively changing how teachers view science is a challenging task (Mulholland & Wallace, 1996; Palmer, 2002).

The goal of this study is to examine whether inquiry-based science content classes might function to help break this cycle by improving preservice teachers' attitudes at a critical juncture before they begin their professional lives as classroom teachers. While there is much research on the positive impact of inquiry on outcomes for K–12 students (Borman, Gamoran, & Bowdon, 2008; Diamond et al., 2014; National Research Council, 2012b), we build on a smaller body of qualitative research regarding the benefits of inquiry instruction in college for preservice elementary teachers (Mulholland & Wallace, 1996; Palmer, 2002). Specifically, we suggest that students can become empowered and enthusiastic about the domain of science through active involvement in the process of inquiry, defined as engaging in the pursuit of scientific questions via data collection, experimentation, exploration, and discussion (National Research Council, 2000).

To examine this issue, we collected data from the *hands-on science* (HoS) undergraduate program at the University of Texas at Austin (Ludwig et al., 2013) to determine whether exposure to these inquiry-based science content courses promoted a change in the science attitudes of a sample of over 200 preservice elementary teachers. In exploring preservice teachers' attitudes, we go beyond the typical singular focus of much research on self-efficacy to instead consider personal attitudes toward science across several dimensions (van Aalderen-Smeets, Walma van der Molen, & Asma, 2012). Additionally, to ensure the robustness of our results, our design utilizes a comparison group of noneducation and nonscience majors enrolled in more traditional lecture-based science courses; our analyses also account for students' social and academic background. Our study offers promising evidence that science content classes in college can be a positive vehicle for changing the attitudes of future elementary teachers, and subsequently has potential implications for the opportunities to learn both cognitive and noncognitive skills that are offered to future generations of elementary science students.

## LITERATURE REVIEW

### Framework: Considering Attitudes Across Multiple Dimensions

When exploring attitudes toward science, it is critical to recognize multiple relevant dimensions. Based on a comprehensive review of prior research on teacher attitudes, and motivated by the lack of substantive clarity and empirical transparency of most prior research, van Aalderen-Smeets et al. (2012) recently advanced a new theoretical framework to provide a cohesive model that captures primary teachers' attitudes toward science. Specifically, they developed a tripartite model of primary teachers' attitudes toward science that distinguishes between three overarching dimensions, each of which is composed of different elements: (1) *perceived control* (which includes elements such as self-efficacy), (2) *affective states* (which includes enjoyment and anxiety), and (3) *cognitive beliefs* (such as perceived relevance). This framework is informed by earlier theoretical models of attitudes (Eagly & Chaiken, 1993), but departs from prior models by considering perceived control (e.g., self-efficacy) as a core dimension, and furthermore defining behavioral intentions as a consequence rather than a component of science attitudes. Their model also calls for researchers to make a clear distinction regarding whether the focus is on teachers' personal attitudes toward science or their professional attitudes toward teaching science, as studies that combine teachers' views of science as a domain with their views on science instruction in their own classroom into one empirical scale blur the object of teachers' attitudes, making substantive interpretation difficult.

The three dimensions of attitudes (whether personal or professional) advanced by van Aalderen-Smeets et al. (2012) are logically related to one another. For example, Bursal and Paznokas(2006) found that teachers with higher levels of self-efficacy had lower anxiety, a finding supported by several other studies (Bleicher, 2007; Palmer, 2002). Yet while related, they nevertheless capture somewhat distinct thoughts and beliefs. For instance, an individual might expect to master an activity or believe that it is useful, but nevertheless find it is unappealing (e.g., flossing their teeth) or even anxietyproducing (e.g., running 10 miles). Therefore, considering elements of all three dimensions is critical to developing a comprehensive picture of teachers' attitudes toward science.

Yet most of the literature about the attitudes of elementary science teachers (either preservice or in-service) focuses on their perceived control in the form of self-efficacy, as there is relatively scant research on either the cognitive beliefs or affective attitudes of teachers (van Aalderen-Smeets et al., 2012). Thus, a key contribution of our study is our consideration of elements of all three dimensions of attitude to more fully capture the complexity of preservice teachers' attitudes toward science.[1] Below we discuss the prior literature on the science attitudes of preservice elementary teachers in more detail, including how such attitudes may influence the outcomes of future students. Because our research questions and subsequent empirical analyses focus on preservice teachers well before they actually enter the elementary classroom, we concentrate on literature on personal attitudes toward science. We then turn to a discussion of why inquiry-based science content classes for preservice teachers have the potential to increase individuals' attitudes across all three dimensions.

---

[1]In this paper, we address all three dimensions of van Aalderen-Smeets et al.'s (2012) theoretical model, but do not discuss (or model) every element within each dimension. For example, while the authors include context dependency as an element that falls under the dimension of perceived control, we do not address this here as it refers to the support that practicing teachers receive from their administrators and therefore is not particularly relevant for a study concerning the personal (rather than professional) attitudes of preservice teachers.

***Perceived Control: Considering Self-Efficacy.*** A key element of the attitudinal dimension of perceived control is self-efficacy, which according to Bandura's (1977, 1982) foundational work is defined as an individual's belief that she can successfully master a situation or deal with an obstacle that arises. An individual's self-efficacy has logical implications for her subsequent behaviors and choices, as she is likely to attempt to avoid those situations or activities where she does not feel she can be efficacious, and persist where she feels confident that she can be successful. While research has demonstrated that in-service elementary teachers exhibit low levels of efficacy in their science teaching (see, for example, Atwater, Gardner, & Kight, 1991; Harlen, 1997; Ramey-Gassert et al., 1996), not surprisingly this pattern is also evident among preservice teachers, who feel less efficacious about their own ability to learn science. For example, Skamp (1991) determined that less than half of the preservice elementary teachers in his study reported having even a fair amount of confidence in their science ability. Similarly, Bleicher's (2007) study of preservice elementary teachers found that participants in a science methods class exhibited low scores on science self-efficacy scales. Low efficacy is often attributed to prior negative educational experiences in science. For example, in a study of five preservice teachers, Mulholland and Wallace (1996) found that their respondents reported very little confidence in their own science abilities and attributed this to negative experiences in their own science schooling as children.

***Affective States: Considering Enjoyment and Anxiety.*** Individuals' affect toward a domain represents another attitudinal dimension. As articulated by van Aalderen-Smeets et al. (2012), the dimension of affect can be further categorized into the positive element of enjoyment and the negative element of anxiety. Beginning with the former, Liang and Gabel (2005) found that most preservice elementary teachers in their study reported that science had never been enjoyable for them. These teachers also indicated that they took science courses only because it was required for their degree program. Smith (2000) and Howes (2002) found similar results. Additionally, the preservice participants in all of these studies attributed their low levels of enjoyment to negative experiences in either (or both) their high school and college science courses, a point to which we will return to later.

Anxiety captures the negative aspect of the affective dimension. Early research by Mallow (1981; also Mallow & Greenburg, 1983) as well as Westerback (1984) define science anxiety as the fear, worry, or apprehension that some individuals experience when presented with the task of learning science. While research on the math anxiety of preservice teachers is extensive (Udo, Ramsey, & Mallow, 2004), and research considering both math anxiety and science anxiety find a strong association between the two (Cady & Rearden, 2007), there is comparatively little research focusing specifically on science anxiety (Bursal & Paznokas, 2006). Yet several studies do provide evidence that preservice elementary teachers report relatively high levels of science anxiety about teaching science, as well as more generalized anxiety about learning science themselves (Cady & Rearden, 2007; Udo et al., 2004; Westerback & Long, 1990).

***Cognitive Beliefs: Considering the Relevance of Science.*** Finally, we discuss research on preservice elementary teachers' beliefs in the relevance of science, a critical element of the attitudinal dimension of cognitive beliefs. The limited extant research on this topic focuses on perceptions of science as useful or relevant for society and for them personally, and finds that preservice elementary teachers report generally favorable views. Specifically, in a study of 200 preservice teachers, Coulson (1992) used a survey instrument that included a personal usefulness science scale and found that on average, respondents indicated moderate levels of agreement. Cobern and Loving's (2002) study at a large Midwestern university also found that on Likert scales measuring the perceived importance of science to all citizens, preservice elementary teachers on average agreed with this sentiment. These findings echo the sentiments of the general population, which generally regard science and technology as useful for making their lives better (Evans & Durant, 1995; Kohut, Keeter, Doherty, & Dimock, 2009). Therefore among the three attitudinal dimensions discussed, promoting preservice teachers' views of the relevance of science is perhaps somewhat less of a pressing problem than promoting both their perceived control in the form of science efficacy and their affective attitudes of enjoyment and anxiety.

## Examining the Impact of Teacher Attitudes

There is a logical connection between teachers' attitudes toward a subject and student outcomes, both in terms of impacting students' opportunities to learn science and their own developing attitudes. First, research indicates that the negative attitudes toward science held by many elementary teachers are likely to result in less coverage of science content and less engaging and effective instruction. Teachers who are not confident in their own science knowledge are likely to worry that they cannot effectively answer students' questions nor keep them engaged with interesting activities (Jarrett, 1999). Consequently, elementary teachers who feel less efficacious and more anxious are likely to try to avoid teaching science and spend less time teaching when they cannot avoid it altogether (Brownlow, Jacobi, & Rogers, 2000; Pine et al., 2006; Ramey-Gassert et al., 1996). Appleton and Kindt (1999) also found evidence that avoidance occurred when teachers did not find science to be as relevant as other subjects such as English or math. One such teacher reported that "If you're running out of time in the week, you think 'Oh I just won't worry about that science activity'" (p. 162).

When teachers do teach science, their negative attitudes can impact their pedagogical practices and subsequently their capacity to reach and engage students. Appleton and Kindt's (1999) study of in-service elementary teachers found that those exhibiting low confidence were less likely to engage their students in hands-on learning. Ramey-Gassert et al. (1996) found that elementary teachers with low science self-efficacy had a minimal desire to engage in professional development activities that could improve their teaching of science. Lack of belief in the usefulness or relevance of science could also logically result in a reduced effort to provide engaging instruction to students.

Additionally, the negative attitudes toward science held by many elementary teachers can result in the socialization of students toward a negative stance toward science. As students look to their teachers as role models and authorities, they begin to mimic and potentially internalize such attitudes as their own (Jussim & Eccles, 1992; McKown & Weinstein, 2002). This can lead to a vicious cycle where negative attitudes such as anxiety and low efficacy are passed from teachers to their students, and therefore from one generation to the next. For example, Beilock et al. (2010) found that elementary teachers' math anxiety influenced students' own attitudes toward math, with girls in particular more likely to evidence declining math attitudes and increasingly gender-stereotyped views of math over the course of the year. This particular study highlights additional concerns regarding gender rolemodeling; as most elementary teachers are female, and a substantial number of them exhibit negative attitudes toward science (as well as math), this could be a strong conduit through which young girls begin to believe that these subjects are less interesting, less important, and overall less-suited for them. Furthermore, as students' own attitudes decline, so does their engagement and motivation to learn, which further impacts their achievement (Eccles, 1994).

In sum, the research literature provides compelling evidence that teachers' negative attitudes can impact students' learning and attitudinal outcomes, and as such are a critical issue that needs to be addressed. To effectively break this cycle, necessitates intervening to change teachers' attitudes before they enter the classroom. In short, the education of preservice teachers is an ideal place to focus.

### Educating Preservice Teachers: Inquiry as a Tool for Improving Attitudes

As mentioned earlier, the negative attitudes of preservice elementary teachers can at least in part be traced back to their own negative experiences with science classes in high school and in college. For example, Liang and Gabel (2005) found that preservice teachers attributed their lack of enjoyment of science to their prior experiences in classes dominated by lectures that necessitated copious amounts of note-taking and memorization. Similarly, Smith (2000) reports that preservice elementary teachers in his study often complained about the boredom of their procedure-based high school and college science courses. Simply put, many preservice teachers have had little exposure to inquiry-based science instruction in their lives as students. We posit that exposure while in college to pedagogy that actively involves them in the process of scientific discovery can ultimately help them to see that science is meaningful, interesting, and accessible, thereby changing their attitudes toward science.

Several studies support the supposition that inquiry classes can promote such changes, in spite of the possibility of some initial student frustration with an approach that deviates from teachers' didactic presentations of the "right" answer (Volkmann, Abell, & Zgagacz, 2005). For example, in a qualitative study of preservice teachers at a large university in the southwest, Kelly (2000) found that after completing an active, inquiry-based science methods course, most participants reported that their interest in science had increased. Kelly (2000) attributes this shift to the use of hands-on explorations and discussions where students came to embody the process of scientific inquiry by formulating and exploring ideas. Similarly, in a study of preservice elementary teachers, Palmer (2002) interviewed four participants who reported that their attitudes toward science had changed from negative to positive due to the excitement of inquiry-based lessons. A study of five teachers by Mulholland and Wallace (1996) reported similar results. Finally, in a quantitative study of 112 preservice elementary teachers, Jarrett (1999) found that an inquiry-based science methods class increased the participants' personal interest in science. The author attributed this change to preservice teachers becoming active agents in the classroom and learning to view science as a process of discovery. Thus, a small body of research finds that inquiry-based pedagogy in science educational methods courses can have a positive impact on preservice elementary teachers' attitudes.

### CURRENT STUDY

Building on the insights of this prior research, we posit that *required science content courses* could also represent a powerful venue for change for college students at the beginning stages of preparing for their careers as future teachers. Specifically, the purpose of this study is to examine whether inquiry-based science content courses promote a change in attitudes toward science among preservice elementary teachers. The courses are part of the HoS undergraduate program, developed at The University of Texas at Austin in the College of Natural Sciences, with cooperation from the College of Education. HoS is required of all students in the elementary education program and covers four semesters of science courses with a curriculum that is composed heavily of the topics that preservice teachers will be expected to teach their students once they become teachers (Ludwig et al., 2013). The design of the HoS program is based on the Physics and Everyday Thinking (Goldberg, 2008) framework that centers around the development of students' physical science understandings through experimentation and follows the example of work from Western Washington University extending this framework to other disciplines (Nelson, 2008). The curriculum is based upon big ideas in science, with specific emphasis on the themes of Matter and Energy, which are integrated across different science disciplines. The course sequence focuses on physics in Semester 1, chemistry and geology in Semester 2, biological systems in Semester 3, and astronomy and earth science in Semester 4.

HoS classes were designed to utilize the essential elements of inquiry-based learning as defined by the influential National Research Council report on inquiry (Forbes, 2011; National Research Council, 2000); specifically, students engage in scientifically oriented questions by collecting, organizing, and analyzing data. From that data they formulate explanations, connect them to scientific knowledge, and subsequently evaluate their explanations in contrast to alternative explanations. Additionally, students share and justify those explanations with others.

The HoS program is best categorized as teacher-directed or guided inquiry (Cuevas, Lee, Hart, & Deaktor, 2005; National Research Council, 2000; Volkmann et al., 2005). Initial questions are posed by the instructor and all activities are carefully designed to present opportunities for students to confront misconceptions, with both topics and skills developed in a structured progression. The instructor does not lecture but probes student knowledge and offers helpful nudges in the right direction, working with students both in small groups and via whole class discussions to construct knowledge and develop explanations (Volkmann et al., 2005). Finally, consistent with inquiry approaches to learning, the course emphasizes big ideas in science while engaging students in hands-on learning and constant exploration and discussion while in groups with their peers (National Research Council, 2000, 2012b).

An example of a typical 2-hour class session is a lesson on sound that begins with the following teacher-generated questions: *How does sound travel through a room? How does sound travel from your classmate to you?* Students then work in groups of three or four to answer these questions and discuss their initial ideas or preconceptions; they subsequently share their ideas with the entire class, so that there is a collective knowledge of alternative ways of thinking about how sound travels. Working in their small groups, students then engage in a series of experiments, using an "airzooka" and candles, a string telephone, and a loudspeaker, that require that they gather data and then generate models based on these data. They are regularly prompted by the instructor to make predictions and connect trends to other previously seen concepts. Finally, students reflect and summarize key ideas and connect them to other contexts by using evidence-based reasoning from the data collected in their experiments. These questions are the basis for whole-class discussions, where students present and justify their ideas to their peers. Students typically end lessons by writing a scientific narrative summarizing how their thinking was revised from their initial ideas, drawing on evidence gained through the experiments and whole class sharing and discussion. For example, a student might share how her initial idea that "sound travels as a wave" has become more sophisticated, discussing the role of vibrations and the transfer of mechanical energy from the source to the receiver.

Our examination of whether the inquiry-based science content courses of the HoS program promote a change in attitudes toward science among preservice elementary teachers is described in detail below. Here, we briefly note that our study contributes to the literature a rigorous quantitative examination of change in attitudes across multiple dimensions among preservice elementary teachers, utilizing a comparison group of students in more traditional science content courses, as well as accounting for differences among students in academic and social background characteristics that may have implications for their attitudes.

## Data and Method

***Analytic Sample.*** Our analytic sample includes 238 preservice elementary teachers who enrolled in the HoS program between Fall 2010 and Spring 2012. Students completed presurveys prior to the start of the first course and postsurveys at the end of the second course in the sequence.[2] Surveys were done online, and were administered during class by members of the research team. Additionally, our sample includes a comparison group composed of 263 nonscience and noneducation majors enrolled in traditional lecture-style undergraduate science courses in Spring 2012. While our design falls shorts of a pure treatment versus control comparison, we chose courses that represent what our sample of preservice elementary teachers would have been required to take in the absence of the HoS program. Therefore, our comparison group includes students who were enrolled in either an introductory-level chemistry or biology class. These students were surveyed at the beginning and end of the semester. As the time period between pre- and postsurveys is longer for HoS students than for non-HoS students, we discuss the implications and limitations of this comparison later.

Administrative records were linked to student surveys, allowing us to examine the demographic and academic background information of students in our sample and to consider how HoS students differed from those in the comparison group. Not surprisingly given the gender composition of the elementary teacher population nationwide, HoS students were overwhelmingly female (95%), compared to 65% of our comparison sample of noneducation and nonscience majors. There were no statistically significant differences between the two groups of students by mother's education; for race/ethnicity, there were significantly fewer students who identified as American Indian among HoS students compared to non-HoS students. Regarding academic background prior to college entry, HoS students had lower SAT math scores than their fellow students taking typical science classes by more than one-third of a standard deviation (see Table 1). To ensure that any differences observed between the two groups in their changes in attitudes over time are not confounded by differences in their background characteristics, our subsequent multivariate models will control for all of these factors.[3]

***Student Attitude Surveys.*** The pre- and postsurveys administered to both groups consisted of 21 items geared toward assessing student attitudes toward science. To examine multiple dimensions of attitudes, we selected items from preexisting surveys to measure

---

[2]At the time of this study, students were only required to take the first two semesters of the total four semester sequence.

[3]Mother's education level is an ordinal variable with the following categories: (1) did not attend high school, (2) attended high school but did not graduate, (3) high school diploma or GED, (4) some college, (5) earned associate's degree, (6) bachelor's degree, (7) graduate or professional degree. Math SAT score and mother's education level were imputed for those students who had missing values. We utilized information on gender, race, high school rank, family income, and father's education to conduct single imputation using Stata. Analyses using list-wise deletion produced similar results.

**TABLE 1**
**Descriptive Statistics**

| | Hands-on Science (HoS) Students | | Non-*HoS* Students | |
|---|---|---|---|---|
| | *N* = 238 | | *N* = 263 | |
| Gender | | | | |
| Female*** | 94.5% | | 65.4% | |
| Race/ethnicity | | | | |
| White (non-Hispanic) | 63.4% | | 60.1% | |
| Black | 2.9% | | 3.4% | |
| Hispanic | 21% | | 19% | |
| Asian | 11.3% | | 14.4% | |
| American Indian** | 0.4% | | 2.7% | |
| Native Hawaiian/Pacific Islander | 0.8% | | 0.4% | |
| Other background characteristics | Mean | SD | Mean | SD |
| SAT math score*** | 574.87 | 75.60 | 607.62 | 73.39 |
| Mother's educational level | 5.10 | 1.60 | 5.07 | 1.69 |

***$p < .001$, **$p < .01$, *$p < .05$, $p < .10$.

confidence (an element of perceived control)[4], enjoyment and anxiety (positive and negative elements of affective states), and relevance (a key element of cognitive beliefs). To measure anxiety, we used the Math Anxiety Rating Scale (Hopko, 2003) and substituted the word science for all references to math. To measure all other attitudes, we selected items developed by the National Center for Education Statistics (www.nces.ed.gov), and used in national surveys, including the Educational Longitudinal Study, the High School Longitudinal Study, and the U.S. component of the Trends in International Mathematics and Science Study. Principal component analyses with promax rotation using the complete set of 21 survey items revealed four factors with Eigenvalues greater than one. We subsequently created separate scales (described below) composed of the individual items that loaded onto each of the four factors. The scales clearly corresponded to the elements of confidence, enjoyment, anxiety, and relevance; using the full sample, the Cronbach's alpha for each scale was .8 or higher.[5]

The *confidence* scale is composed of four items that gauge students' level of confidence when engaging in scientific activities. The items were as follows: "I have always done well in science," "Science is not one of my strengths" (reverse coded), "I am confident that I can understand the most difficult material presented in my science textbooks," and "Science has always been one of my best subjects." The five items in the *enjoyment* scale capture students' level of positive affect for science and included "I enjoy learning science," "I look forward to going to science courses," "Science is fun," "I like science," and "Science is boring" (reversecoded for inclusion in the scale). Categories of response were the same as those for the confidence scale.

[4]Because the questions ask students to reflect more on their assessment of their current success in science, we refer to this scale as measuring confidence rather than self-efficacy for future success. However, prior research has consistently noted a very strong correspondence between indicators of self-confidence and self-efficacy (Wigfield & Eccles, 2000).

[5]To further test the reliability of the four scales, we calculated Cronbach's alphas separately for (a) treatment and reference groups; (b) first year cohort and second year cohorts (for treatment group), and (c) survey responses at time 1 and time 2 (for different cohorts and for different treatment groups). In all, we calculated alpha reliabilities for each of our four scales for 15 different subsamples with remarkably consistent results ranging from a low of .77 to a high of .88. (exploratory factor analyses using each of these groups also yielded the same four factor model).

For each of the eight items in the *anxiety* scale, students were asked to indicate how much the situation made them feel anxious or worried. Similar to the other measures, the responses ranged from (1) not at all anxious to (5) very much anxious or worried. The items included "Looking through the pages in a science text," "Thinking about an upcoming science test one day before the test," "Reading and interpreting a scientific graph, chart, or illustration," "Taking an exam in a science course," "Watching and listening to a teacher explaining a scientific concept or phenomena," "Waiting to get a science test returned in which you expected to do well," "Walking on campus and thinking about a science course," and "Being given a 'pop' quiz in science class."

Finally, the four items making up the *relevance* scale include questions that focused on students' perception of the meaning or usefulness of science in their daily lives. The items are as follows: "The subject of science is not very relevant to most people," "It is not important for most people to understand science," "I think learning science will help me in my daily life," and "Science is important to me personally." Students indicated the extent to which they agreed or disagreed with the statements with possible responses ranging from strongly agree (1) to strongly disagree (5).

## ANALYSES AND RESULTS

To examine whether inquiry-based science content courses promote a positive change in attitudes toward science for preservice elementary teachers, we begin by comparing the means on the pre- and postmeasures of our four attitudinal scales for HoS students. Results of paired *t*-tests indicate a statistically significant improvement over time for each scale. Specifically, for confidence, the mean increased from 2.61 to 2.98 ($p < .001$), reflecting a 0.37 point increase or almost half of a standard deviation change from pre to post. HoS students' science enjoyment increased by about a fourth of point (and about a third of a standard deviation), from the presurvey mean of 3.24 to a postsurvey mean of 3.51 ($p < .001$). The largest change was observed for the anxiety scale, where the average decreased (meaning that students became less anxious over time) from a presurvey mean of 3.11 to a postsurvey mean of 2.63 ($p < .001$), a difference of almost half of a point and approximately two-thirds of a standard deviation. Finally, as discussed earlier, students view science as a relevant domain, as evidenced by a relatively high presurvey mean of 3.66 on the utility scale. Nevertheless, they slightly increased their views of the relevance of science after taking HoS courses ($p < .05$) to a postsurvey mean of 3.74, a change of about a tenth of a standard deviation.

Subsequently, we turn to an examination of how the changes we observe for HoS students compare to changes in attitudes toward science among students taking more traditional lecture-based science content courses. Here, we utilize multilevel mixed-effects models, an extension of regression analysis that is similar to a mixed-design analysis of variance and appropriate when data are nested. For this study, repeated measures of attitudes are nested within individuals with time treated as a random effect (Rabe-Hesketh & Skrondal, 2008).The goal of this analysis was to determine whether the change in different dimensions of science attitudes observed for HoS students was similar or different than that observed for non-HoS students while controlling for students' background characteristics (which is particularly important as our two groups of students differed by gender and math SAT scores, both factors which likely predict attitudes).

**TABLE 2**
**Regression Analyses Predicting Attitudes to Science**[a]

| | Model 1 Confidence | Model 2 Enjoyment | Model 3 Anxiety | Model 4 Relevance |
|---|---|---|---|---|
| Hands-on science (HoS) students (ref = non-HoS students) | −0.391*** | −0.163* | 0.106 | −0.051 |
| | (0.071) | (0.072) | (0.066) | (0.059) |
| Time (change from pre- to postsurvey) | −0.127** | −0.131*** | 0.074∼ | −0.116** |
| | (0.039) | (0.039) | (0.041) | (0.036) |
| Time × *HoS* | 0.500*** | 0.391*** | −0.553*** | 0.200*** |
| | (0.056) | (0.056) | (0.060) | (0.052) |
| Female | −0.402*** | −0.395*** | 0.311*** | 0.046 |
| | (0.085) | (0.084) | (0.076) | (0.069) |
| Race/ethnicity (ref = white) | | | | |
| Black | 0.078 | −0.069 | 0.018 | −0.159 |
| | (0.188) | (0.185) | (0.168) | (0.149) |
| Hispanic | −0.136 | 0.054 | 0.055 | −0.023 |
| | (0.093) | (0.091) | (0.083) | (0.073) |
| Asian | −0.314** | −0.110 | 0.232** | −0.002 |
| | (0.098) | (0.097) | (0.088) | (0.078) |
| American Indian/Alaska native | −0.591* | −0.181 | 0.287 | 0.112 |
| | (0.251) | (0.249) | (0.224) | (0.204) |
| Native Hawaiian/other Pacific Islander | 0.635 | 0.380 | −0.524 | 0.537∼ |
| | (0.418) | (0.407) | (0.374) | (0.322) |
| SAT math score | 0.002*** | −0.000 | −0.002*** | 0.000 |
| | (0.000) | (0.000) | (0.000) | (0.000) |
| Mother's education level | −0.038∼ | −0.013 | 0.029 | −0.003 |
| | (0.022) | (0.022) | (0.020) | (0.017) |
| Constant | 2.680*** | 4.179*** | 3.748*** | 3.762*** |
| | (0.320) | (0.315) | (0.288) | (0.255) |

[a]Coefficients calculated from a multilevel regression model in Stata where observations of attitudes are nested within individuals.
***$p < .001$, **$p < .01$, *$p < .05$, ∼$p < .1$; standard errors in parentheses; $n = 501$.

Table 2 displays the results of separate analyses for each dependent variable. The first row displays the coefficient comparing HoS and non-HoS students on the presurvey (or time 1) for each attitudinal dimension. The second row displays the average change over time between the pre- and postsurvey, while the third row displays the interaction between student group (HoS or non-HoS) and time. The change in attitudes from pre- to postsurvey for HoS students is calculated as the sum of the main effect of time and the interaction term, while change for non-HoS students is captured by the main effect of time only. To simplify the presentation of results, we include figures for each attitudinal outcome (Figures 1–4) that display the changes over time for each group, adjusted for the social and academic characteristics discussed above.[6]

[6]Figures 1–4 display the adjusted pre and post means for each group. These are calculated using a postestimation command in Stata where all other variables in the model (other than student group, time, and the interaction) are set to the mean (or alternatively to the mode for categorical variables).

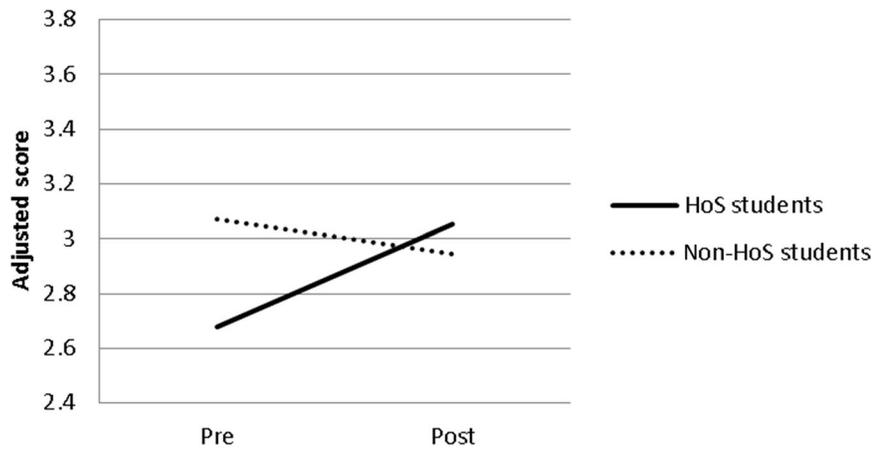

**Figure 1.** Confidence.

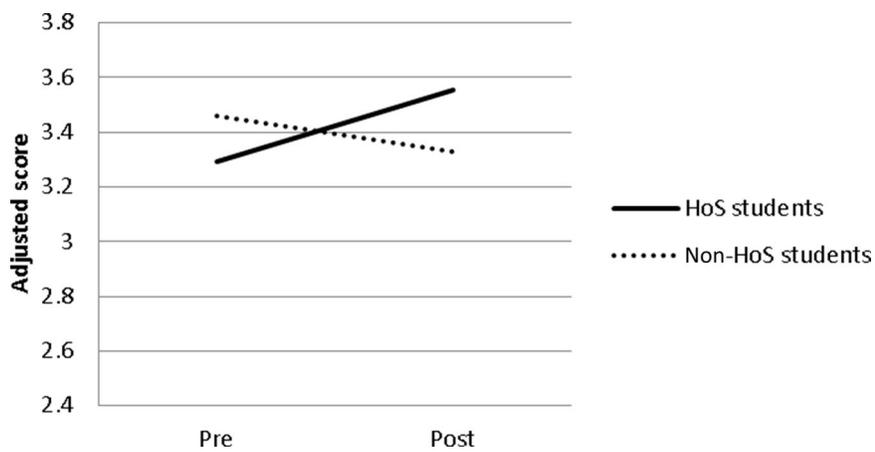

**Figure 2.** Enjoyment.

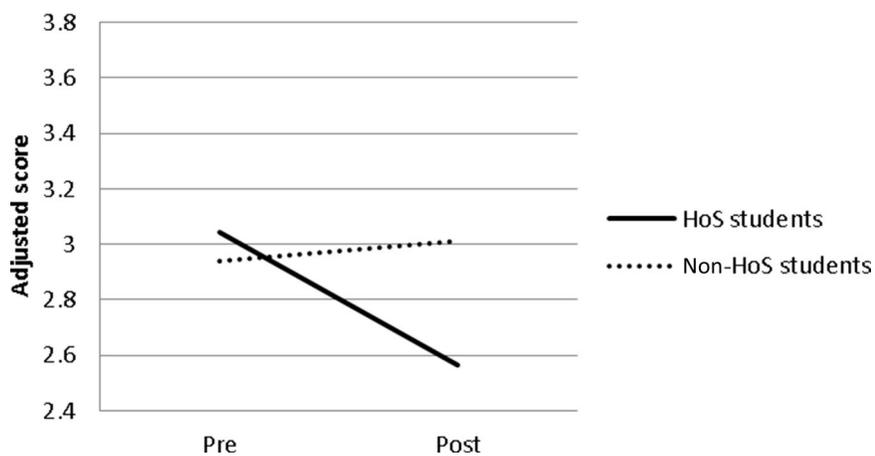

**Figure 3.** Anxiety.

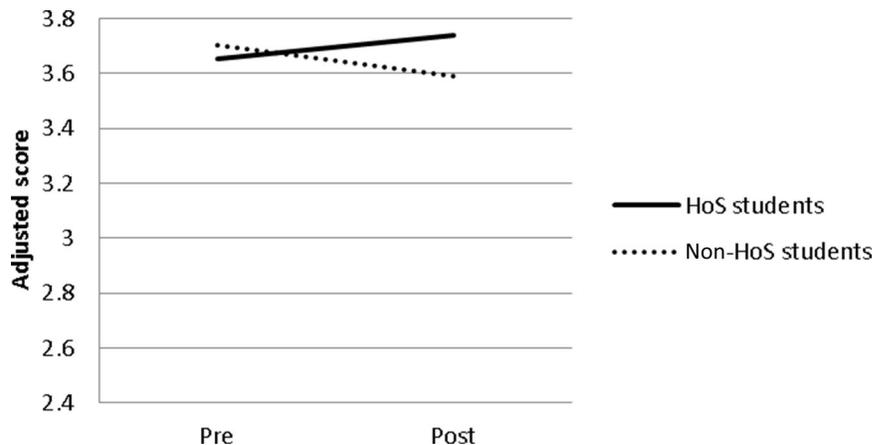

**Figure 4.** Relevance.

Beginning with the first column predicting *confidence,* the results reveal that compared to their peers in traditional science classes, HoS students reported significantly lower science confidence on the presurvey (−.391***). The coefficient for time is negative and significant, yet the interaction between time and student group is positive and significant, indicating that HoS students increased their confidence over time relative to non-HoS students. To help clarify the patterns for the two groups, Figure 1 displays the trends for each group. Here, we see clearly that changes in attitudes occurred for both groups in opposite directions. While HoS students initially had lower confidence than their non-HoS peers, they significantly increased their confidence over time. In contrast, non-HoS students significantly decreased their science confidence (−.127***), such that their confidence was slightly lower than non-HoS students on the postsurvey.

Returning to Table 2, we see a similar pattern when predicting change in science *enjoyment*. HoS students initially reported significantly lower levels of enjoyment than their peers in more traditional classes (−.163**). However, once again we see a negative main effect of time but a positive and significant interaction between student group and time, indicating opposite directions of change for each group. As displayed graphically in Figure 2, HoS students significantly increase their enjoyment over time, while on average non-HoS students report a decrease in their affect toward science (−.131***), and end their course reporting lower enjoyment than HoS students.

Regarding changes in *anxiety,* we note that HoS and non-HoS students do not differ significantly on the presurvey. The main effect of time is positive and borderline significant, yet the interaction term reveals a statistically significant difference between the two groups' average change in anxiety. Figure 3 clearly shows the marked decrease in anxiety from the pre- to the postsurvey for HoS students. For their non-HoS peers, however, the figure shows a slight increase in anxiety (∼.074).

Finally, Table 2 displays the results for models predicting changes in attitudes toward the relevance of science. HoS and non-HoS students do not differ significantly on the presurvey. But once again the interaction term reveals a statistically significant difference between groups in change over time, and the sign of the coefficient is positive in contrast to negative main effect of time. Figure 4 displays the disordinal patterns for the groups. HoS students' views of relevance increase a small amount from the pre- to the postsurvey, whereas their peers' perceptions of the relevance of science decreases (−.116**).

Finally, while the main focus of our analyses was to assess differences between HoS and non-HoS students regarding changes in their science attitudes, our multivariate analyses revealed patterns consistent with prior research, namely that females are significantly less confident in their science ability and report significantly less enjoyment and more science anxiety than their male peers (Correll, 2001; Eccles, 1994). Our models also indicate some evidence of racial/ethnic differences in attitudes (see the models predicting confidence and anxiety), as well as differences by prior math achievement, as those with higher scores on the math portion of the SAT report significantly higher levels of confidence in their science ability and less anxiety. Given the differential distribution of HoS and non-HoS students on several of these covariates (see Table 1), including them in our models generally decreased the size of the difference between groups on the presurvey (e.g., the more male composition of the non-HoS group helped to account for their initially stronger math confidence), and also revealed larger differences between the groups on the postsurvey than could be detected in simpler models that did not account for these factors.

## DISCUSSION AND CONCLUSION

The goal of this study was to focus on future elementary teachers' views toward science at a point when they are still explicitly occupying the role of a learner, before they are tasked with assuming the role of a science teacher. Specifically, we investigated whether enrollment in HoS, a program of inquiry-based science content courses, promoted a favorable change in the science attitudes of a sample of over 200 preservice elementary teachers. Building on the theoretical framework advanced by van Aalderen-Smeets and colleagues (2012), our study examined changes in attitudes on multiple dimensions, and also utilized a comparison group of noneducation/nonscience majors to help contextualize the changes observed for our focal sample of elementary preservice teachers.

Data analyses reveal a remarkably consistent and positive story for HoS students; students significantly changed their views toward science from the pre- to the postsurvey, such that after participating in inquiry-based content courses they reported more confidence in their skills as science learners, more enjoyment and less anxiety toward science, and perceived it as more relevant. Conversely, patterns for those in the comparison group revealed a decline in favorable attitudes toward science after enrolling in a traditional, lecture-based content course. Importantly, these results are independent of differences between HoS and non-HoS students (e.g., gender and SAT math score) that are associated with attitudes; therefore, our analyses indicates that it was differences in the courses that the two groups of students enrolled in, rather than characteristics of the individual students themselves, that led to changes in attitudes.

Our study has several likely implications for the future classrooms of preservice teachers, as prior research suggests that teachers with negative views toward science may both socialize their young students to develop similar views, as well as offer less science instruction in class due to a desire to avoid the subject (Bursal & Paznokas, 2006; Jarrett, 1999). Thus, to the extent that HoS students have more positive attitudes toward science as a result of inquiry-based instruction, we have positively intervened to disrupt the vicious cycle of elementary school teachers passing their negative views of science onto the next generation. Instead, future elementary students could ultimately be the beneficiaries, having teachers who are favorably inclined and excited about teaching science and therefore spend more time and focus on it.

Additionally, we suggest that our results have potential implications for gender equity in the classroom, particularly because, while all observed changes for HoS students were in a favorable direction, the largest changes were for decreasing anxiety. Recent research by Beilock et al. (2010) offered evidence that the math anxiety of female elementary teachers had a negative impact on their female students in particular. The authors argue that due to the inclination for children to more strongly connect to adults of the same gender as role models, young girls in the classroom were more susceptible to teachers' math anxiety, and consequently exhibited more negative attitudes of their own as well as lower math achievement. Their study provides powerful evidence that teacher role-modeling can be a key factor that leads to the development of the gender gap in math as early as elementary school (Beilock et al., 2010). Such a pattern can be logically extended to science anxiety; thus by intervening to decrease the anxiety of (predominantly female) preservice elementary teachers, more young girls could have the opportunity to interact with a positive female role model, thereby thwarting emerging gender disparities in children's views and performance (Eccles, 1994).

Our study also has potential ramifications for the type of instruction and classrooms experienced by future elementary students. As research suggests that preservice teachers tend to teach their students in ways similar to how they were taught (Gess-Newsome & Lederman, 1999), programs such as HoS can serve as a powerful model for implementing inquiry in their own classrooms, and thus ultimately contribute to greater uptake of recommended science reform. The Next Generation Science Standards call for elementary teachers to engage their students in inquiry-based practices, as well as emphasize disciplinary core ideas in the classroom (National Research Council, 2012b). The HoS classes offer preservice teachers the critical opportunity to develop an understanding of and real experience with inquiry-based teaching and learning that is consistent with these standards. We concur with Volkmann et al. (2005), who argue that "if learning through inquiry is to become a reality in today's schools, then university science courses must model inquiry so that pre-service teachers may experience it" (p. 867). Programs such as HoS have the power to do exactly this, and thereby help break the cycle of teacher-centered didactic instruction.

While the primary focus of this study is the educational experiences of preservice teachers during college and the subsequent implications for future elementary classrooms, our study also speaks to the need to improve undergraduate science education more broadly. A recent meta-analysis of student academic performance in STEM undergraduate courses provides evidence that traditional lecture formats lead to higher failure rates and lower achievement when compared to classes that are more constructivist based (Freeman et al., 2014 ). Our study focuses on attitudinal rather than performance outcomes, and in doing so heeds recent calls by the National Research Council to examine a more comprehensive range of student outcomes at the postsecondary level (National Research Council, 2012). Specifically, we find that, even after adjusting for differences in social and academic background between our two groups (preservice education students and noneducation/nonscience students), enrollment in traditional lecture classes has the opposite effect of enrolling in inquiry-based content classes. We suggest that the decline in confidence, affect, and utility, as well as a slight increase in anxiety that we observed for students in traditional lecture-based science content classes, are important consequences of their relatively low engagement in the classroom, and perhaps linked to patterns of lower performance documented elsewhere (Freeman et al., 2014; Seymour & Hewitt, 1997).

As with any study, ours has limitations. First, we note a lack of parallel time frames for our focal HoS students (two semesters between pre- and postsurveys) and our comparison group (one semester between pre- and postsurveys). Therefore, it is possible that part of the positive change in attitudes observed for HoS students is due to the longer exposure period; while we cannot dismiss this possibility entirely we are nevertheless skeptical that a shorter survey window for HoS students would have substantively changed our findings regarding opposite directions of change for the two groups. Indeed, we did collect postsurveys for a small subsample of HoS students at the end of the first semester, and the results, although slightly weaker in magnitude, were statistically significant and in the same positive direction as the full sample of HoS students included here.

Additionally, we did not collect data to assess what features of these inquiry-based classes students found most favorable; for instance, it could be that the significant amount of time spent on group work was a particularly influential factor leading to their change in attitudes (Park Rogers & Abell, 2008). We think this is an important area for future research to address, to better ascertain which aspects of inquiry-based classrooms are most effective at promoting favorable shifts on different dimensions of science attitudes. More long-term studies are also needed to assess whether and how the potential implications we discuss above come to fruition and make a difference for elementary students in the science classroom.

Finally, it is important to point out that designing and implementing inquiry-based science content courses that depart from the typical lecture-based format of most postsecondary instruction is certainly not without its challenges and difficulties (Allen & Tanner, 2005; Armbruster, Johnson, & Weiss, 2009). One such obstacle is convincing science instructors and university administrators of the benefits of inquiry-based instruction for their students. Our study contributes to the small number of studies that offer robust empirical evidence on this topic (Seymour, 2002), and in doing so offers additional support to the call to implement inquiry-based science instruction at all levels and for all students.

This research was supported by a grant from the National Science Foundation (NSF DUE 0942943, PI: Sacha Kopp), and a grant from the Eunice Kennedy Shriver National Institute of Health and Child Development (5 R24 HD042849) awarded to the Population Research Center at The University of Texas at Austin. Opinions reflect those of the authors and do not necessarily reflect those of the granting agencies.